# Topic-aware Large Language Models for Summarizing the Lived Healthcare Experiences Described in Health Stories

Maneesh Bilalpur, MS[1], Megan Hamm, PhD[2], Young Ji Lee, PhD [3,4], Natasha Norman[2], Kathleen M. McTigue, MD[2*], Yanshan Wang, PhD[1,4,5*]

[1]Intelligent Systems Program, University of Pittsburgh, Pittsburgh, Pennsylvania, USA;

[2]Department of Medicine, University of Pittsburgh, Pittsburgh, Pennsylvania, USA;

[3]School of Nursing, University of Pittsburgh, Pittsburgh, Pennsylvania, USA; [4]Department of Biomedical Informatics, University of Pittsburgh, Pittsburgh, Pennsylvania, USA;

[5]Department of Health Information Management, University of Pittsburgh, Pittsburgh, Pennsylvania, USA

* Co-corresponding authors: Kathleen Mctigue: KMM34@pitt.edu; Yanshan Wang: yanshan.wang@pitt.edu

**ABSTRACT**

**Objective:** Storytelling is a powerful form of communication and may provide insights into factors contributing to gaps in healthcare outcomes. To determine whether Large Language Models (LLMs) can identify potential underlying factors and avenues for intervention, we performed topic-aware hierarchical summarization of narratives from African American (AA) storytellers.

**Materials and Methods:** Fifty transcribed stories of AA experiences were used to identify topics in their experience using the Latent Dirichlet Allocation (LDA) technique. Stories about a given topic were summarized using an open-source LLM-based hierarchical summarization approach. Topic summaries were generated by summarizing across story summaries for each story that addressed a given topic. Generated topic summaries were rated for fabrication, accuracy, comprehensiveness, and usefulness by the GPT-4 model, and the model's reliability was validated against the original story summaries by two domain experts.

**Results:** Twenty-six topics were identified in the fifty AA stories. The GPT-4 ratings suggest that topic summaries were free from fabrication, highly accurate, comprehensive, and useful. The reliability of GPT ratings compared to expert assessments showed moderate to high agreement.

**Discussion**: Our approach identified AA experience-relevant topics such as health behaviors, interactions with medical team members, caregiving and symptom management, among others. Such insights could help researchers identify potential factors and interventions by learning from unstructured narratives in an efficient manner – leveraging the communicative power of storytelling.

**Conclusion:** The use of LDA and LLMs to identify and summarize the experience of AA individuals suggests a variety of possible avenues for health research and possible clinical improvements to support patients and caregivers, thereby ultimately improving health outcomes.

**INTRODUCTION**

Storytelling is a narrative style that captures information beyond objective facts by catering to beliefs and emotions of both the narrator and the audience. Sharing lived experiences with others through stories can overcome the limitations due to extreme objectivity in alternatives like the didactic communication[1,2]. Storytelling can successfully share health knowledge[1-4], potentially providing insight into how illness shapes people's lives, promote behavior change, and suggesting avenues for improving health care[5-11]. Such stories – sharing the topics deemed most important by the storyteller – can suggest innovative avenues for interventions to improve the healthcare ecosystem. Hence there is a need for the health informatics community to investigate narratives of healthcare experiences.

Despite the efficacy of narratives in healthcare, their usage poses certain challenges. Information stored as oral recordings is less accessible and searchable than structured data. Furthermore, moving from a single anecdote to the derivation of common themes can be expensive and time-consuming. The use of Natural Language Processing (NLP) techniques may help to overcome these challenges. Prior studies have examined structured feedback from patients[12], free-text feedback from social media[13], and Electronic Health Records (EHR)[14] to identify themes or topics using NLP techniques. Often, topic detection has been paired with sentiment analysis to infer an individual's sentiments towards the topic. Such solutions offer automated approaches for understanding limitations in healthcare environments, efficacy of treatment approaches, and overall experience in patient care to provide tailored solutions[12]. However, existing literature has largely focused on structured, written-language data, with relatively limited exploration of unstructured sources such as spoken-dialog narratives.

Spoken-dialog narratives present a unique set of challenges; it is syntactically and semantically different from written language[15] and is often long and tends to cover multiple topics in a single dialog. Under such conditions, topic-specific sentiment analysis is a challenge due to a lack of defined topic boundaries in the data. Summarization approaches overcome this limitation. They are also more informative as they provide fine-grained analysis when compared to single sentiment labels from conventional sentiment analysis. Recent rapid expansion in applications of LLMs has shown that they excel at summarizing documents across domains (including healthcare[16]) and demonstrated their efficacy at understanding spoken language[17]. Thus, they offer an effective solution towards summarizing spoken dialog. Despite their ability to comprehend, LLMs are limited by their input context length. For example, the popular open-source LLM, the LLaMA 3.2 has an input context length of 128k tokens. However, topics tend to cover multiple stories whose length often exceeds the fixed context length of LLMs. In addition, long-form inputs (commonly seen in spoken-dialog narrations) are a challenge to handle and run the risk of forgetting intermediate portions of the document[18]. On the other hand, classical methods are not susceptible to such limitations, especially during inference. In this work, we evaluate the comprehension capabilities of LLMs in long-form dialog understanding by leveraging advantages of classical methods in topic detection to build a topic-aware summarization model for understanding spoken-dialog narratives about healthcare experiences.

In this work, we analyze spoken narratives of experiences (referred to as "stories") from the African American (AA) population to identify underlying topics and summarize issues raised about the storytellers' lived healthcare experience. The stories are a subset of those stored in the MyPaTH Story Booth archive [termed Story Booth]. We focus on stories by AA storytellers because, compared to white individuals in the United States, AAs have worse health outcomes and are less likely to receive health care services[19-26], despite numerous efforts to close such gaps[26-28]. Additionally, storytelling may be a particularly appealing avenue for understanding AAs' health experiences since the community has a strong oral tradition[29, 30, 31], and may also foster trust-building[24].

We focus on understanding individual experiences in the healthcare ecosystem. Storytellers include patients, caregivers, and healthcare professionals. Topic detection was performed over transcriptions of the long-form spoken-dialog dataset using Latent Dirichlet Analysis (LDA). The LDA topics were augmented with topic labels for clinical interpretability using LLMs. Furthermore, we introduce topic-aware summarization to generate topic summaries through summaries of individual stories by topic [termed topic story summary], extracting the nature of the healthcare experience by each topic from stories, as depicted in the topic story summaries, using LLM-based summarization.

**BACKGROUND**

Free-text patient comments have been previously studied using a supervised approach for topic detection[32]. Doing-Harris et al.[32] performed manual annotations for topics and sentiment labels to develop automated solutions for topic detection and found appointment access and wait, empathy, explanation, friendliness, practice environment, and overall experience as frequent topics in patient feedback. Similarly, another work[14] studied EHR records to identify goals-of-care in patient care. Topic detection was used to identify suicide profiles from 300,000 decedents[33]. They found that suicide profiles broadly covered 5 classes: including *mental health and substance problems*, *mental health*

*problems*, *crisis, alcohol-related, and intimate partner problems*, *physical health problems*, and *polysubstance problems*. Furthermore, they found demographic shifts in the suicide profiles in the profiles with time showing an evolving landscape of healthcare needs. A recent review[21] of NLP techniques to understand patient-experience feedback compared literature in terms of ML techniques (supervised vs. unsupervised) and data collection approaches (social media vs. structured surveys). Among supervised approaches, Naïve Bayes classifier was best performing while the unsupervised LDA model was the popular alternative for topic detection. The ease of data collection through structured written language has led to its popularity in understanding healthcare experience feedback. Limited work explored spoken dialog feedback. A recent work studied supervised models for topic detection in patient-doctor conversations[34]. They used turn-level manual annotations for gold-standard reference. They varied the input context length for the supervised models and found that topic detection improves with an increase in input context length.

We focus on the quality of clinical interactions, its outcome, and the broader impact on daily-to-day life of patients, providers, and caregivers to identify potential causes and areas of healthcare disparity. Our work leverages fine-grain details in participants' conversational style data about healthcare experience. We believe that such details offer a nuanced understanding of problems in the healthcare experience of marginalized communities. We achieve this using a topic-aware hierarchical summarization approach of healthcare feedback. In the following sections, we describe the dataset, techniques for topic detection and hierarchical summarization, and present our findings about healthcare experience from the generated summaries.

**DATASET**

The MyPaTH Story Booth[35] archive is a collection of individual experiences, told from a patient, caregiver, or healthcare provider perspective. Developed as community engagement infrastructure for the PaTH Clinical Research Network, the archive includes over 1,500 stories related to experiences with illness, efforts to maintain health, and interactions with health care systems[36]. Story recordings were conducted in-person or by telephone. Participants can opt to share an unstructured story or to answer prompts selected from a pre-approved list. They are asked to limit their stories to twenty minutes in length or less. Participants are adults recruited from research registries and through advertisements in community and clinical venues. Informed consent is obtained from all participants; the

University of Pittsburgh's IRB and data storage and computation guidelines were followed (IRB protocol numbers: STUDY19020307, STUDY20110315).

The dataset used in this work is limited to 1,120 stories – the full set collected by University of Pittsburgh staff and fully processed at the outset of this study. From this set, 50 stories from AA participants were randomly selected for summarization and analysis. We used OpenAI's Whisper[37] model for diarization (speaker detection and transcription) of the audio recordings. Due to the limited involvement of the interviewers, we focused only on the participants' statements. We validated the Whisper transcriptions against manual transcriptions from the 50 validation stories to observe a Levenshtein distance[38] of 6.2% for character-level changes (insertions, deletions, or substitutions). Manual inspection found that the diarization quality is satisfactory and the differences in labeling conventions between manual and automatic Whisper approaches to be the major source of diarization errors.

**METHODS**

Our method for topic-aware summarization of long-form healthcare experiences combines classical methods of topic detection with recent advances in LLMs through a multi-step process. First, we perform topic detection using the LDA[39] approach. The identified topics are then labelled for interpretability using LLMs and topic story summaries were generated. Finally, topic labels and story-topic distributions from the LDA are leveraged for topic-aware summarization using a hierarchical summarization approach powered by LLMs. Individual steps are further described below. Figure 1 shows various components of the proposed approach.

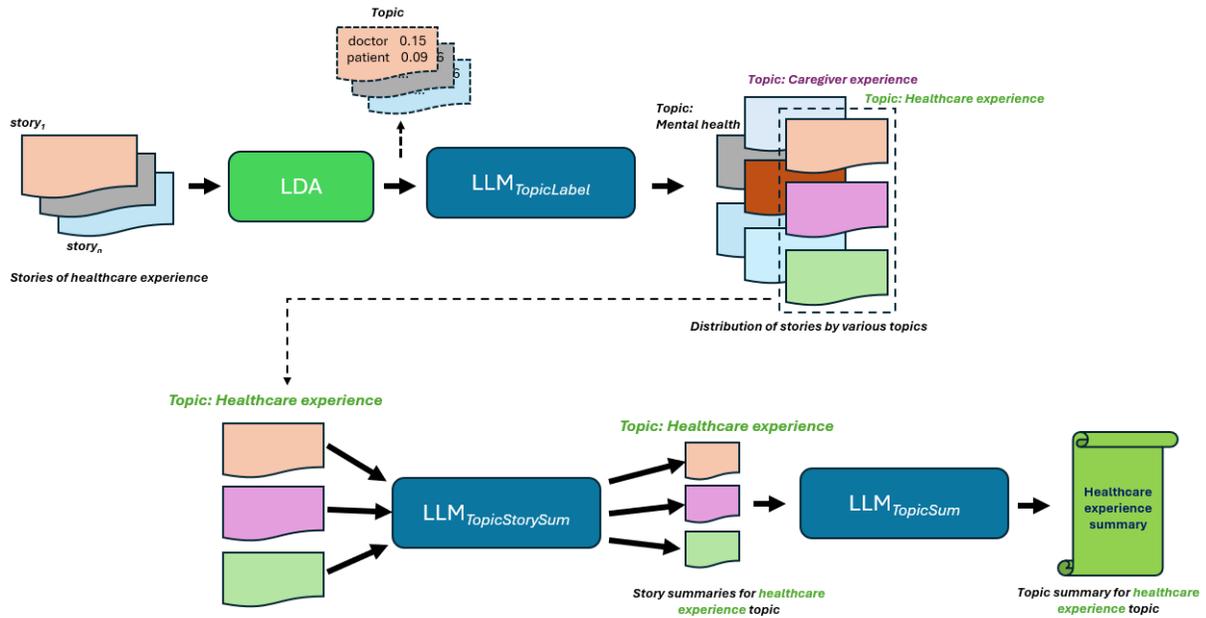

**Figure 1.** Various steps in the proposed approach for Topic-aware Large-Language Model for Long-form Spoken-dialog Summarization.

**Topic detection and labelling**

To understand recurring topics from the healthcare experience stories, it is necessary to identify the topics across stories. Topic modeling approaches like LDA are often the go-to solutions for such problems and widely used in studying healthcare experience as discussed in the related work. LDA is a probabilistic topic modeling to capture the likelihood of topics across documents (here, stories) and simultaneously the distribution of words in each topic. We use LDA for detecting topics across all 1,120 stories from the dataset. The topic detection model was tuned by varying the number of topics between 50 to 1,000 in steps of 50. The optimal number of topics were identified using perplexity criterion on our validation set.

The word distributions in each topic offer a naïve understanding of topics. They present the likelihood of word(s) in each topic but are often hard to interpret. They may also contain contradictory, ambiguous, or unrelated words under the same topic. This is more prevalent in our dataset considering the participant perspectives include patients, caregivers, and healthcare professionals, each offering a variety of perspectives which makes understanding the conversational themes more challenging. To overcome this limitation in interpretation, we labelled word clouds of topics using a pretrained LLaMA-3.1 model[40]. The LLaMA is an open-source LLM with competitive performance against closed-source alternatives such as GPT-4[41] and GPT-3.5 Turbo[42]. For all LLM experiments, we use the 70 B parameter model due to its long context length of 128k tokens. We optimize the inference time using the *llama.cpp*[43] framework with 4-bit quantization and GPT-Generated Unified Format (GGUF) encoding.

Given the challenging nature of the task, we derived labels for each topic in the context of each story. The LLM input consisted of a story and the word list from the topic-word distribution *(see topic labeling prompt template in figure 2)*. The words were ranked in the order of their likelihood in the topic as determined by the LDA. Topic labels derived in the context of a story tend to overfit to a specific story. Moreover, across multiple stories the topic label was found to be paraphrases of one another. To establish consistency of topic labels across all relevant stories, the most frequent label of the topic across all stories was chosen as the final topic label.

**Hierarchical summarization**

**Algorithm 1.** Topic-aware hierarchical summarization using Large-Language Models

---

**Require:** Dataset $D = \{s_1, s_2, ..., s_n\}$ of $n$ stories. $D_{valid}$ is the validation set.
**Require:** $LLM_{TopicLabel}, LLM_{TopicStorySum}, LLM_{TopicSum}$ be the LLM configured for topic labeling, summarizing a story given topic, and summarizing a topic given one or more story summaries respectively.
**Ensure:**
    $TopicStory, WordTopic \leftarrow LDA(D)$    ▷ train LDA for likelihood for topic-story and word-topic distributions
    $TopicStory_{valid}, WordTopic_{valid} \leftarrow LDA(D_{valid})$    ▷ predict likelihood for topic-story and word-topic distributions on $D_{valid}$
    Let $Topics_{valid}$ be the set of topics where $TopicStory_{valid} \leq TopicProbThresh$    ▷ TopicProbThresh = 0.05
    **for** topic $t$ in $Topics_{valid}$ **do**
        $TopicLabels \leftarrow list()$    ▷ label for topic $t$ inferred from each story $s$
        **for** story $s$ in $D_{valid}$ **do**
            $TopicLabels \leftarrow LLM_{TopicLabel}(story_s, WordTopic_{w,t})$
        **end for**
        $TopicLabel_t \leftarrow mode(TopicLabels)$    ▷ most common label across stories is used as the final topic label
        **for** story $s$ in $D_{valid}$ **do**
            $TopicStorySum_{t,s} \leftarrow LLM_{TopicStorySumm}(s, TopicLabel_t)$
        **end for**
        $TopicSum_t \leftarrow LLM_{TopicSum}(TopicStorySum_{t,\cdot}, TopicLabel_t)$
    **end for**
    **return** $TopicSum$

---

*Algorithm 1* presents a step-by-step overview of our approach. We leverage the LDA-identified topics to summarize the stories from the validation set. To overcome the input context length limitations, a hierarchical summarization approach was used. Our hierarchical summarization approach involves two steps. In the first step, topic summarization for each story is performed *(see topic story summarization prompt template in figure 2)*. This greatly consolidates the input and overcomes the limitation of input context length for long-form summarization with LLMs. Following this, individual topic story summaries are further summarized to generate a holistic summary of all the stories under the topic *(see topic summarization prompt template in figure 2)*. The hierarchical summarization also offers interpretability through tracing of elements of the holistic summary.

```
[TOPIC LABELING PROMPT]

[SYSTEM]Given the participant experience dealing with the healthcare system,
identify the topic labels that fit the experience based on the given list of
topic words. Each list should correspond to only one topic label. Given
output in <TOPIC LABEL>: LIST OF WORDS format. No additional text is required
in the output.
[USER] Participant experience: <STORY>. List of topic words: <WORDS IN STORY>
________________________________________________________________________
[TOPIC STORY SUMMARIZATION PROMPT]

[SYSTEM]Summarize the experience of participant dealing with the healthcare
system along the given topic. Use participant(s) own words during
summarization and do not paraphrase.
[USER] Participant(s) experience: <EXPERIENCE> Topic label: <TOPIC LABEL>
________________________________________________________________________
[TOPIC SUMMARIZATION PROMPT]

[SYSTEM] Generate a holistic summary experience from individual
participant(s) summaries about dealing with the healthcare system along the
given topic. Note that all summaries should fit and not deviate from the
topic label. Do not include any additional information apart from that
present in the input. Do not paraphrase.
[USER] Participant(s) summaries: <PARTICIPANT STORY SUMMARIES>. Topic label:
<TOPIC LABEL>
```

**Figure 2.** Prompts used with LLMs for topic labeling, and topic story summarization and topic summarization for hierarchical summarization.

*Evaluation*

A strong evaluation of generative models is critical to understanding their limitations. Conventional evaluation metrics such as ROUGE[44] only account for co-occurrence similarity between generated and reference text. Hence in the context of LLMs these similarity-based metrics make little sense. LLM-generated content is also prone to a distinct set of challenges such as hallucinations and inadvertent safety limitations that are not studied with such conventional metrics. To account for these shortcomings, we use the multidimensional evaluation framework, QUEST[45] to evaluate the generated summaries. The evaluation dimensions for this summarization task include comprehensiveness, fabrication, accuracy, and usefulness.

Human evaluation is often expensive and time-consuming, so we assessed the LLM's ability to generate topic summaries using the LLM-as-a-judge approach. To eliminate the self-enhancement bias in LLMs, we used the GPT-4 Turbo as our judge. GPT-4 was found to correlate well with human raters and less susceptible to position-bias[46]. Individual topic summaries together with the respective topic story summaries were used as input to rate the four QUEST dimensions on a five-point Likert scale (see Table S1 in the Supplementary Material for the definition of our Likert scale). We quantify the reliability of the GPT responses against two domain experts as human raters. Raters were assigned four topics (chronic pain management, medical treatment, caregiving experience, and health concerns) to rate the QUEST dimensions on the same Likert scale as the GPT model. Discrepancies in Likert values between the two human raters were discussed and adjudicated by those raters to provide a final rating which was compared to the GPT responses. We use Bennett's S-score[47] with quadratic weighting to calculate agreement between the raters and the GPT responses.

**RESULTS**

The LDA for topic detection yielded 150 optimal topics across the training set. To further investigate the topics present in the validation set, we tuned the topic-likelihood threshold for stories to identify topics present in each story. A large threshold results in too few topics while a small threshold leads to too many. We struck a balance for the choice threshold using an intuitive approach. The threshold was set to accommodate at least one topic for each validation set story. This resulted in 40 topics identified in the validation set, see *figure 3* for the word clouds for some sample topics.

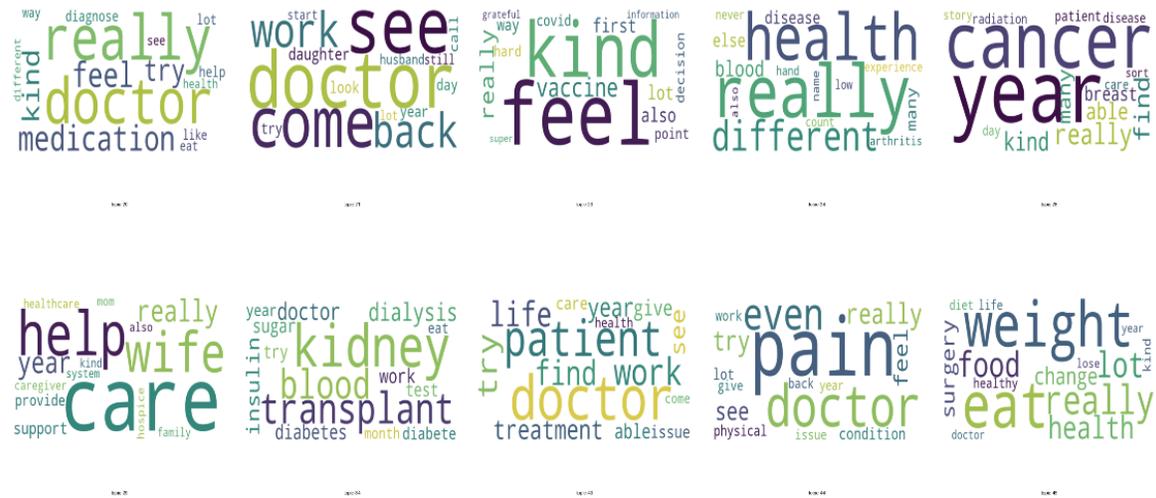

**Figure 3.** Sample word cloud representation of 40 topics identified by LDA in the validation set (zoom to see details).

Topic labels for the LDA topics are necessary for our topic-aware summarization approach. Such labels also improve topic interpretability for domain-experts. As mentioned in the methods section, our topic labeling approach using LLM resulted in 26 topic labels across all the stories (see Table 1 for topic labels) and the remaining 14 topics were omitted due to the inherent difficulty in labeling topics based on a distribution of words and redundancy among the labels. These topic labels were then used in the summarization step.

**Table 1.** Topic labels derived from the LLM. Numbers in parenthesis are the number of stories that address the topic.

| LLM-derived topic labels | |
|---|---|
| Healthcare Experience (38) | Healthcare System (26) |
| Healthy Eating (13) | Caregiving Experience (11) |
| Doctor-Patient Relationship (11) | Symptom Management (11) |
| Cancer Treatment (10) | Chronic Pain Management (10) |
| Hospital Experience (9) | Surgical Experience (7) |
| Doctor Experience (6) | Medical Treatment (6) |
| Personal Experience (5) | Health Concerns (3) |
| Disease Diagnosis (3) | Mental Health (3) |
| Healthcare (3) | Diagnosis (2) |
| Caregiving (2) | Diabetes Management (1) |
| Cancer Diagnosis (1) | Health Challenges (1) |
| Emotional and Physical Impact (1) | Medical Condition (1) |
| Heart Health (1) | |
| Symptoms and Hospitalization (1) | |

The topic labels together with the corresponding stories are used to generate topic summaries through the intermediate topic story summaries. The results of hierarchical summarization for two sample topics are tabulated in Table 2. Table S3 in Supplementary Material lists summaries from the corresponding topics. We find the summaries for each topic are comprehensive, coherent, and aligned with the topic label derived based on the LDA topic-word distribution.

**Table 2.** Sample summaries for two different topics from our hierarchical summarization approach. See Table S3 in Supplementary Material for summaries from corresponding topics.

| Topic label | Topic summary |
|---|---|
| Chronic Pain Management | Participants dealing with chronic pain management reported various challenges and experiences. Many **struggled with inadequate pain management, including ineffective treatments, lack of effective pain medication, and hesitation from healthcare providers** to treat their pain. Some participants felt that their pain was not taken seriously, leading to frustration and further suffering. **Others found alternative methods, such as CBD oil, yoga, and herbs, to be helpful** in managing their pain. A few participants benefited from working with supportive doctors, pain clinics, and transportation programs that helped them access necessary care. Despite these efforts, **many participants continued to experience chronic pain, numbness, and debilitating symptoms, affecting their daily lives** and relationships. |
| | Caregivers face numerous challenges, including **lack of support, financial struggles, and emotional toll**. Many participants expressed **the importance of compassion, patience, and affection** when caring for loved ones, particularly those with dementia or Alzheimer's. Some participants **had to navigate complex healthcare systems, advocate for their loved ones' needs, and deal with the emotional impact** of caregiving. Others |

| Caregiver Experience | **appreciated the support of family members, healthcare providers, and online support groups**. Several participants highlighted the **need for respite care, financial assistance, and guidance for caregivers**. Some also emphasized the **importance of acknowledging stress and seeking help**. Additionally, participants noted that **healthcare providers should advise patients on healthy living and provide resources for caregivers**. Overall, caregiving experiences varied, but many participants shared the common theme of needing more support and resources to effectively care for their loved ones." |
|---|---|

**Evaluation**

**Table 3:** Bennett's S-score agreement between raters and GPT averaged over four topics after adjudication. The score ranges from -1 to 1 where -1, 0, and 1 indicate perfect disagreement, chance-level agreement, and perfect agreement respectively. S (A, B) denotes agreement between raters A and B.

| QUEST dimension | S (R1, R2) | S (GPT, R1) | S (GPT, R2) | $\sum_i$ S (GPT, $R_i$) / 2 |
|---|---|---|---|---|
| **Fabrication** | 0.94 | 0.94 | 1.00 | 0.97 |
| **Accuracy** | 0.81 | 0.62 | 0.69 | 0.65 |
| **Comprehensiveness** | 0.94 | 0.94 | 0.87 | 0.91 |
| **Usefulness** | 1.00 | 0.75 | 0.75 | 0.75 |

We evaluate the generated summaries with topic story summaries to perform human evaluation. Table 3 presents the inter-rater and GPT-rater agreements from our evaluation approach. We notice a moderate to high agreement between the GPT responses and raters. The level of agreement is particularly strong for fabrication and comprehensiveness of generated summaries. This suggests that GPT evaluation of generated summaries overlap with human raters and can be used reliably to evaluate our topic-aware hierarchical summarization model. Table S2 in the

Supplementary Material presents the GPT responses for all topics present in the validation set. The accuracy of generated summaries is consistently good for most topics except *diagnosis* where the generated summary likely changed the interpretation of the topic because of one substantive point in it. The usefulness of the summaries was found to be mostly high indicating that our approach is promising to use in place of manually identifying themes in the story and summarizing them as defined by our Likert scale for evaluation.

In addition to the ratings from experts, their free-form comments suggest that our transcription and summarization is error-prone. Errors such as incorrect dosage units when referring to medication in-take ("I take *30 kilos* of medication a day. I used to take *50 kilos* a day") were due to difficulties in challenges in our Whisper-based transcription approach where "30 pills" was incorrectly transcribed as "30 kilos". Lexical inconsistencies when referring to a single story ("*Many* participants expressed the importance of compassion, patience…" was actually expressed by only person), deviating from the topic of interest, and a lack of cohesion in topic with fewer stories can be observed as well.

**DISCUSSION**

Our approach of using automated transcription, topic modeling, and summarization introduces a computational method to analyzing narratives of healthcare experiences. Some common topics that were identified from our dataset include healthcare experiences, healthy eating, caregiving, doctor-patient relationships, symptom management, cancer treatment, and chronic pain management. Through our summarization approach, we expand the scope of understanding healthcare experiences beyond the conventional coarse sentiment analysis[13] to offer details related to clinical interactions, treatment efficacy and their outcomes from individual narratives. Our hierarchical summarization approach is particularly useful for extracting such insights from large narrative datasets. The summaries for all 26 topics identified from the narratives can be found in the Supplementary Materials in Table S4.

An important component of our proposed approach is topic modeling using the LDA. Among 150 topics identified in the training set, 40 were present in the validation set. Interestingly, only 26 topics can be labeled for topic names using the LLM. Despite the modest number of topics, redundancy among the topics is noticeable. Topics such as *healthcare experience* and *healthcare system, caregiving experience and caregiving, and doctor-patient relationship and doctor experience* are closely related. We would like to highlight that such redundancy in topics is an expected outcome of LDA and the choice of hyperparameters, addressing it remains an active area of research in NLP[49]. While LDA is a popular choice for topic modeling, it assumes a Dirichlet distribution of topics in a document. We believe models such as the Probabilistic Latent Semantic Indexing (pLSI)[50] may result in different topics. We would also like to highlight that in addition to the probabilistic generative approaches like the LDA and pLSI, the NLP community has also begun exploring LLMs for topic modeling[51,52].

The summarization step leverages the superior summarization capabilities of LLMs to offer additional insights into the narratives. Our hierarchical summarization overcomes the challenge of limited input context of the LLMs by generating individual topic story summaries followed by topic summaries for corresponding stories. Our evaluation framework addresses the limitations of conventional metrics for summarization such as the ROUGE using a multidimensional approach derived from the QUEST framework. We found that the GPT-4 Turbo evaluations concur with human evaluations and the summaries our approach provides summaries that are comprehensive, accurate, free of fabrication, and useful.

Human evaluation is often expensive and time-consuming. Here, we found that LLMs can help to summarize story contents, identifying several relevant insights and key topics in an efficient and timely manner. However, trade-offs are involved, such as considerable topic redundancy and uncertainty regarding accuracy in summary details. Maintaining a human element in the evaluation process can help to identify and address such concerns.

**LIMITATIONS AND FUTURE WORK**

This work also has some limitations. The automatic transcription was evaluated against only a small set of human transcriptions. Accounting for diverse transcription errors due to differences in speaker accents, linguistics, and environment variables through a larger dataset remains a future work. Recent work[51,52] studied the efficacy of LLM and prompting techniques on topic modeling to find that LLMs can identify topics and offer explainability. We believe this direction may help address the topic redundancy and lack of explainability limitations of conventional topic models like the LDA. The hierarchical summarization uses a 70 B parameter LLaMA LLM due to their generalization and emergent capabilities[53], however, the rapidly evolving LLM landscape warrants a comparison between alternative choices such as the Qwen[54], and DeepSeek[55] to help identify the optimal model for summarization. The GPT-based approach for summarization evaluation was limited to comparing topic summaries with individual story-topic summaries rather than comparing the topic summaries against the original transcripts. This assumes that the generated story-topic summaries are sufficiently comprehensive, accurate, and free from any fabrication. Future work shall incorporate human evaluation of generated topic summaries through comparison with the original stories from each topic to better assess summary accuracy. The quantitative measures of agreement in human evaluation should be interpreted keeping the limited focus on four topics in perspective. The small dataset (50 stories) limits our topic detection and summarization outcomes. However, it reflects a design choice put in place to enable a future analysis directly comparing LLM and traditional qualitative evaluations. Future work shall consider larger datasets collected from different geographical locations and healthcare infrastructures for a better understanding of experiences. Pipeline-based approaches like our topic detection followed by hierarchical summarization are brittle due to error propagation between successive steps. End-to-end approaches overcome this limitation; we believe research to explore end-to-end approaches by integrating topic modeling together with summarization is another exciting direction for future research.

**CONCLUSIONS**

This work contributes to the healthcare informatics domain using NLP techniques (topic detection and hierarchical summarization) to offer an understanding of narratives of healthcare experiences beyond the conventional sentiment analysis approach that is popular in existing literature on healthcare experiences. We leverage both traditional and modern NLP techniques, such as the LDA and LLMs, to achieve the same. Using widely accepted quantitative metrics for reliability, we demonstrate that crucial steps in the proposed approach (such as transcription of audio recordings of the narratives and evaluation of summaries) can be performed with sufficient reliability using recent advances in speech processing and LLMs. We believe this provides an opportunity to attract interest from the community and accelerate research using narratives towards improving healthcare experience and health equity using computational methods.

**ACKNOWLEDGEMENTS**

This work is a part of the Story Booth project coordinated by the University of Pittsburgh and funded through Patient-Centered Outcomes Research Institute (PCORI) Awards (RI-PITT-01-PS8; RI-PITT-01-PS1). The views presented in this paper are solely the responsibility of the authors and do not necessarily represent the views of the PCORI, its Board of Governors or Methodology Committee.

**Author contributions**

MB, MH, KMM, and YW: conceptualized the study, conducted data analysis, wrote the manuscript; YJL, NN: conducted data analysis, wrote the manuscript.

**Funding**

This work was supported by the Patient-Centered Outcomes Research Institute (PCORI) Awards (RI-PITT-01-PS8; RI-PITT-01-PS1).

**Conflicts of interest**

The authors declare that they have no conflicts of interest related to this publication.

**Data availability**

The data used in this study are approved for research by IRB and currently not publicly available.

# SUPPLEMENTARY MATERIAL

**Table S1:** Likert-scale definition used for QUEST dimensions used in the GPT-based and human evaluation of topic summaries.

| QUEST dimension and definition | Likert-scale definition |
|---|---|
| **Fabrication** (The summary contains made-up information or data that is not covered in the story. This includes any plausible but non-existent facts**)** | 1. Wholly fabricated.<br><br>2. Largely fabricated.<br><br>3. One substantive point fabricated – i.e., the fabrication changes the meaning/interpretation of the topic of interest.<br><br>4. One minor detail fabricated – i.e., fabrication is present but does not change the meaning/interpretation of the topic of interest.<br><br>5. No made-up information. |
| **Accuracy** (The summary is factually correct, precise, and free from any errors) | 1. Wholly inaccurate.<br><br>2. Largely inaccurate.<br><br>3. One substantive point is inaccurate (i.e., inaccuracy changes the interpretation of the topic of interest).<br><br>4. One minor detail is inaccurate (i.e., inaccuracy is present but does not change the interpretation of the topic of interest).<br><br>5. No inaccuracies in the summary. |
| **Comprehensiveness** (The summary covers all critical themes (i.e., patient concerns about healthcare experience) discussed in the story. It offers a completely comprehensive overview of the story along with sufficient details) | 1. Topic summary covers none of the critical themes from the story summaries.<br><br>2. Topic summary is missing most of the critical themes from the story summaries.<br><br>3. One substantive theme is missing from the summary (which shifts the interpretation of the topic of interest).<br><br>4. One minor theme is missing from the summary (which does not impact interpretation of the topic of interest).<br><br>5. All critical themes are covered in the topic summary. |
| **Usefulness** (The summary is useful. It can be reliably used in place of manually identifying key themes in the story and then summarizing them) | 1. Topic summary cannot be used at all in place of manually identifying key themes from the story summaries.<br><br>2. Topic summary is largely not useful for identifying key themes from the story summaries.<br><br>3. Topic summary has a moderate degree of usefulness for identifying key themes from the story summaries.<br><br>4. Topic summary is mostly useful for identifying key themes from the story summaries.<br><br>5. The topic summary is extremely useful for identifying key themes from the story summaries. |

**Table S2:** GPT-4-Turbo responses to evaluating topic summaries on a 5-point Likert scale. The Likert-scale definition can be found in Table S1.

| Topic | Fabrication | Accuracy | Comprehensiveness | Usefulness |
|---|---|---|---|---|
| Doctor Experience | 5 | 5 | 5 | 5 |
| Medical Condition | 5 | 5 | 5 | 5 |
| Healthcare Experience | 5 | 5 | 5 | 5 |
| Chronic Pain Management | 5 | 4 | 4 | 5 |
| Hospital Experience | 5 | 5 | 5 | 5 |
| Surgical Experience | 5 | 5 | 5 | 5 |
| Caregiving Experience | 5 | 5 | 5 | 5 |
| Healthcare System | 5 | 5 | 5 | 5 |
| Symptom Management | 5 | 5 | 4 | 5 |
| Medical Treatment | 5 | 5 | 4 | 4 |
| Health Challenges | 5 | 5 | 4 | 4 |
| Healthy Eating | 5 | 5 | 5 | 5 |
| Diabetes Management | 5 | 5 | 5 | 5 |
| Doctor-Patient Relationship | 5 | 4 | 4 | 5 |
| Health concerns | 5 | 5 | 5 | 5 |
| Cancer Treatment | 5 | 5 | 4 | 5 |
| Caregiving | 5 | 5 | 4 | 5 |
| Personal Experience | 5 | 5 | 4 | 4 |
| Diagnosis | 5 | 3 | 4 | 4 |
| Healthcare | 5 | 5 | 4 | 4 |
| Disease Diagnosis | 5 | 5 | 5 | 5 |
| Mental Health | 5 | 5 | 5 | 5 |
| Cancer Diagnosis | 5 | 5 | 5 | 5 |
| Heart Health | 5 | 5 | 5 | 5 |
| Symptoms and Hospitalization | 5 | 5 | 5 | 5 |
| **Emotional and Physical Impact** | 5 | 5 | 5 | 5 |

**Table S3:** Topic story summaries used to generate topic summaries for samples used in Table 2. Each paragraph is a separate story summary for the topic.

| Topic label | Topic story summary |
|---|---|
| Chronic Pain Management | I'm still dealing with chronic pain and numbness in my hand after the injury. I've seen a pain doctor and had nerve blockers, but the pain persists. I'm frustrated with the lack of effective pain management and the need for repeated surgeries and procedures. <br><br> The participant struggled with chronic pain due to endometriosis and experienced debilitating symptoms, including heavy bleeding, clots, and nausea. They felt that their pain was not taken seriously by some healthcare providers and were not offered adequate pain management options. <br><br> I have been using CBD oil to manage my chronic pain as conventional over-the-counter treatments don't work for me, and I don't want a pain prescription. On really bad days, I use an ultra-sized tampon or a menstrual cup to help manage my symptoms. <br><br> I was in a lot of pain for a long time, with sharp pains in my stomach and leg. I had to try different pain meds, but they didn't work. I was always in bed and lost my drive. After surgery, I felt better, but I had to learn how to manage my stress and not let it affect my relationships with my kids. |

The participant's aunt suffered from chronic pain due to her diabetes and amputations, but was not provided with adequate pain management, leading to further suffering and decline in her health.

I've been dealing with an injury from 20 years ago that's causing me chronic pain. Despite this, I've been sent to physical therapy, which I don't think will fix the underlying problem. I feel like the healthcare system is trying to avoid giving me the treatment I need, specifically an operation, and instead is trying to manage my pain with temporary solutions.

I have chronic pain and I take 30 kilos of medication a day. I used to take 50 kilos a day, but I've been able to reduce my medication. I have a chronic pain doctor who helps me manage my pain. I also have a transportation program that helps me get to my doctor's appointments, which is free.

The participant struggled with chronic pain management, particularly with regards to their arm and hand. They experienced numbness, tingling, and pain, which made everyday activities challenging. They tried various treatments, including medication, physical therapy, and hand therapy. They also participated in a pain clinic, which helped them learn coping mechanisms and understand how pain affects behavior. They found that the pain clinic was a turning point in their management of chronic pain and that it helped them develop a better understanding of their condition.

I've learned to take care of myself and put myself in positions where I'm not stressing about unnecessary things. I've also had to relearn my body and figure out what triggers my lupus. My doctors have been helpful in managing my pain, and I've had infusions and other treatments to maintain my condition.

| | |
|---|---|
| | Managing chronic pain is a challenge. I've had to deal with hesitation from healthcare providers to treat my pain, which can lead to longer hospital stays and more severe pain. However, I've found that working with a supportive doctor and using holistic approaches like yoga and herbs can help manage my pain. |
| Caregiver Experience | The participant's family members, including their mother and grandmother, took on caregiving roles for their niece and grandmother, respectively. The participant expressed concerns about the lack of support for caregivers, including respite care and financial assistance, and the impact on their quality of life. <br><br> The participant's experience with caregiving was shaped by her young age and stressful circumstances during her pregnancy. She had to navigate abusive relationships, financial struggles, and poor eating habits, which she believes may have impacted her son's health. Despite these challenges, she was grateful for the support of her family and a particular doctor who took the time to ask about her lifestyle and well-being. She learned the importance of acknowledging stress and seeking help, and wishes that she had received more guidance and support as a young mother. <br><br> I don't currently have a caregiver, but I try to stay independent and take care of myself. I also try to support others who are caregivers or patients, and I participate in online support groups to connect with others who share similar experiences. <br><br> The participant shares their experience of being a caretaker to their mother who was diagnosed with dementia Alzheimer's. They emphasize the importance of being compassionate, patient, and affectionate, and accepting the patient's reality. They also highlight the need to involve the patient in activities, allow them to maintain their independence, and show love and support. The participant notes that it's essential to be |

hands-on, listen to the patient, and validate their feelings, even if it means hearing repeated stories. They also stress the importance of not showing anger or frustration, even when the patient becomes aggressive or accusatory.

The participant's experience as a caregiver was emotionally challenging, and they felt that the caregiving system failed to provide adequate support and resources. They had to navigate complex healthcare systems and advocate for their aunt's needs, but ultimately felt that they were not equipped to provide the level of care she required.

The participant had a challenging experience with aftercare, as they were not able to get a caregiver due to insurance limitations. They had to rely on family members and church members for support, and eventually had to find their own way to manage their care. They suggest that healthcare providers should advise patients on healthy living and provide resources for caregivers.

The participant became a caregiver for their father when he was diagnosed with prostate cancer. They had to navigate the healthcare system to get him the care he needed and had to fight to get him tested for the BRCA gene. They also had to deal with the emotional toll of caring for a loved one and eventually had to make the decision to put him in hospice care.

The participant's sister played a significant caregiving role, driving to the hospital and providing support during the recovery process. The participant also appreciated the hospital's arrangement for a nurse to visit their home, ensuring their safety and well-being as they lived alone.

| | The participant's experience as a certified nursing assistant (CNA) is highlighted, where they took care of elderly residents in nursing homes, assisting with daily living activities such as bathing, feeding, and dressing. They also shared a memorable experience of caring for an AIDS patient who was initially resistant to care but eventually warmed up to them. |
|---|---|
| | I took care of my mother for many years, managing her medical appointments and caring for her needs. It was a challenging experience, especially as she became more depressed and withdrawn. I also had help from my family, including my husband and children, when I was recovering from my car accident in 1965. |
| | The participant's adoptive mother was often absent and neglectful, leaving them and their disabled older brother alone for long stretches of time. The participant had to rely on a charity organization and state-funded counselors for support, but felt that their concerns were not taken seriously. |

**Table S4:** Generated summaries for each of the 26 topics identified from the 50 stories of healthcare experience from African American individuals.

| Topic | Topic summary |
|---|---|

| | | |
|---|---|---|
| | Healthy Eating | Many participants emphasized the importance of healthy eating in managing their health conditions, such as pre-diabetes, type 2 diabetes, hypertension, cancer, and sickle cell disease. However, some participants faced challenges in making healthy food choices, including lack of understanding, limited access to healthy food options, and difficulty changing old eating habits. Some participants had to research and figure out healthy eating habits on their own, while others worked with healthcare professionals, such as nurse practitioners, nutritional counselors, and dietitians, to develop personalized eating plans. Common themes included the importance of eating a balanced diet, avoiding processed and sugary foods, and staying hydrated. Some participants also noted the importance of considering individual lifestyle and food preferences when making healthy eating choices. Overall, participants recognized the significance of healthy eating in managing their health conditions and improving their overall well-being. |
| | Chronic Pain Management | Participants dealing with chronic pain management reported various challenges and experiences. Many struggled with inadequate pain management, including ineffective treatments, lack of effective pain medication, and hesitation from healthcare providers to treat their pain. Some participants felt that their pain was not taken seriously, leading to frustration and further suffering. Others found alternative methods, such as CBD oil, yoga, and herbs, to be helpful in managing their pain. A few participants benefited from working with supportive doctors, pain clinics, and transportation programs that helped them access necessary care. Despite these efforts, many participants continued to experience chronic pain, numbness, and debilitating symptoms, affecting their daily lives and relationships. |
| | Doctor Experience | Some participants had negative experiences with doctors, feeling dismissed, talked down to, and disrespected. Doctors were perceived as arrogant, |

|  | uncaring, and unwilling to consider alternative explanations or provide second opinions. In contrast, other participants had positive experiences with doctors who were supportive, listened to their concerns, and provided alternatives and explanations. These doctors were described as wonderful, rock stars, and understanding, taking the time to get to know the participants and their goals. |
|---|---|
| Emotional and Physical Impact | The participant experienced a prolonged period of pain and stress, compounded by the emotional toll of their mother's death and caregiving responsibilities. They continue to suffer from aching and numbness in their leg, necessitating ongoing pain management with medication. |
| Hospital Experience | Participants had varied experiences with hospitals, ranging from disappointment and frustration to relief and gratitude. Some participants felt ignored, like a nuisance, or moved through like cattle, while others appreciated empathetic and supportive healthcare providers. Efficient intake and diagnosis were noted by some, while others experienced challenges with communication and disconnection from healthcare providers. Traumatic experiences, including life-threatening episodes and prolonged stays in critical care, were also reported. However, some participants were grateful for the care they received and appreciated the efforts of hospital staff. Regular hospital visits for ongoing treatments, such as blood transfusions, were also mentioned, highlighting the emotional toll of diagnosis and treatment. |
| Disease Diagnosis | Participants' experiences with disease diagnosis highlight the challenges and importance of seeking professional help. One participant with early-onset Alzheimer's and breast cancer emphasizes the need for proactive health management. Another participant with cardiac sarcoidosis experienced a delayed diagnosis, taking five years to confirm, and wished for earlier |

|  | diagnosis to better understand and manage their care. A third participant with MDS initially felt shock and denial but eventually became an advocate for themselves and others after coming to terms with their condition. |
| --- | --- |
| Cancer Diagnosis | Dealing with the healthcare system after a cancer diagnosis can be challenging, particularly when insurance coverage is disputed. One individual's experience involved a recent job change and new insurance, which initially refused to cover their cancer treatment, citing a pre-existing condition despite the diagnosis occurring just two weeks after starting the job. This led to a prolonged fight with the insurance company to secure coverage for necessary medical procedures. |
| Caregiving | Caregiving experiences were marked by feelings of unpreparedness and insufficient support. Participants took on caregiving roles for family members with serious health conditions, such as end-stage renal failure, and struggled to balance caregiving responsibilities with work and personal life. Despite receiving some help from home health agencies and family members, participants faced difficulties in providing 24-hour care and managing increasing care needs. The caregiving system was perceived as failing to provide adequate support, leading to feelings of regret and frustration when participants were unable to continue caring for their loved ones. |
| Symptom Management | Managing symptoms is a daily challenge for many participants, who struggle with chronic pain, heavy bleeding, nausea, and other issues. Some feel that their healthcare providers do not take their symptoms seriously, leading to frustration and a need for self-advocacy. Participants have found various ways to manage their symptoms, including medication, self-care, and lifestyle changes. Some have learned to plan their day around their medication and make adjustments to their diet to minimize symptoms. |

|  | | |
|---|---|---|
|  | Others have found alternative treatments, such as photo light therapy, and have learned to prioritize stress management and self-care. Despite the challenges, many participants have found ways to persevere and take care of themselves, often with the support of family, doctors, and other healthcare professionals. | |
| Cancer Treatment | **Treatment Approaches and Outcomes**<br><br>* Some participants were able to manage their cancer through surgery, exercise, and early detection, without requiring chemotherapy, radiation, or medication.<br> * Others underwent various treatments, including chemotherapy, radiation, mastectomy, and reconstructive surgery, with some experiencing side effects and complications.<br> * One participant also incorporated alternative therapies, such as acupuncture and Chinese herbs, into their treatment plan.<br> * One participant underwent standard treatment for colon cancer, including chemotherapy and genetic testing, and took preventive measures to reduce their risk of developing other types of cancer.<br><br>**Decision-Making and Patient Autonomy**<br><br>* Some participants reported making informed decisions about their treatment plans, including opting for mastectomy or seeking second opinions.<br> * One Participant was proud of their decision to opt for a mastectomy instead of chemotherapy and radiation, given their HIV-positive status.<br> * One Participant felt that they were not given the option to have genetic testing for the BRCA gene early on, which may have changed their | |

|  | | treatment plan. **Emotional and Psychological Impact** * Some participants emphasized the importance of staying positive, focusing on what they can do, and managing their emotional well-being during and after treatment. * One participant highlighted the emotional impact of being diagnosed with a life-threatening condition, even if it is not cancer. **Importance of Early Detection and Patient Engagement** * Some participants stressed the importance of early detection and managing one's own health in achieving successful treatment outcomes. * Participant suggested that patients should listen to their doctors, take notes, and ask questions to ensure they understand their treatment plan. |
|---|---|
| Diagnosis | Lung cancer diagnosis was unexpected for both participants, with no prior symptoms. Routine doctor's visits and chest x-rays revealed tumors on the upper left lobe of the lung. One participant's diagnosis was prompted by a mention of struggling to reach certain notes while singing, while the other had a routine check-up. Doctors' confidence and thorough explanations helped alleviate initial fear. |
| Health Challenges | Dealing with the healthcare system is a significant challenge due to the complexity of multiple health conditions, including thalassemia minor, diabetes, heart disease, arthritis, scleritis, trigger fingers, tendonitis, corporal tunnel, high blood pressure, asthma, vocal cord dysphonia, and thyroid issues, which collectively impact daily life. |

| | | |
|---|---|---|
| | Medical Treatment | Participants experienced mixed results with medical treatment, with some encountering misdiagnosis, delayed treatment, and unhelpful guidance. Others found relief with healthcare providers who took a holistic approach, listened attentively, and used evidence-based methods. Challenges included traveling for specialized treatment, dealing with medical errors, and affording expensive medication. Some participants faced difficulties in finding effective treatments for chronic conditions like PCOS, endometriosis, and rheumatoid arthritis. |
| | Doctor-Patient Relationship | Participants had varied experiences with their doctors, ranging from positive to negative. Some appreciated doctors who were approachable, willing to simplify medical terms, and listened to their concerns. Others had negative experiences with doctors who were rushed, condescending, dismissive, or unwilling to listen. Some participants valued the importance of a good rapport with their doctor, citing the need for respect, clear communication, and a willingness to work together. A few participants appreciated doctors who were knowledgeable about specific communities, such as the LGBTQ+ community. Some participants also highlighted the importance of being able to ask questions and receive clear explanations about their condition and treatment options. Overall, participants emphasized the need for doctors to be empathetic, communicative, and responsive to their needs. |
| | Heart Health | Managing heart health involves a complex and challenging experience, requiring learning to manage the condition through medication and lifestyle changes, emphasizing the importance of listening to one's body and communicating with doctors for optimal care. |
| | Healthcare | Participants have reported negative experiences with the healthcare system, citing issues with government-assisted healthcare, such as long wait times, packed waiting rooms, and doctors who are not aware of patients' |

| | |
|---|---|
| | backgrounds and needs. Healthcare professionals have also been criticized for not respecting patients' time, rushing appointments, and lacking communication skills. Additionally, some participants have experienced insensitive treatment from healthcare providers, including being asked insensitive questions, being dismissed, and not being listened to, resulting in inadequate care. |
| Surgical Experience | Participants had varied experiences with surgical procedures, ranging from traumatic to positive. Some participants experienced anxiety, pain, and complications, such as infections, implant issues, and unhealed wounds. Others had successful surgeries with quick recoveries. Support from healthcare providers was inconsistent, with some participants receiving empathetic care and others feeling dismissed or unsupported. |
| Diabetes Management | Dealing with the healthcare system for diabetes management involves struggling to find the right support until meeting a specialized nurse practitioner who provides necessary guidance. This support includes understanding the diagnosis, developing a treatment plan, and making lifestyle changes that improve health. Additionally, it involves learning the importance of self-advocacy and seeking specialist care. Ultimately, it leads to empowerment to educate others about effective diabetes management and the value of supportive care. |
| Health Concerns | Managing multiple health issues is a common challenge, as seen in the experiences of these participants. They face difficulties in keeping track of medications and appointments while dealing with various health concerns such as heart problems, knee problems, low blood pressure, cancer, thyroid problems, high blood pressure, arthritis, and MDS. These conditions significantly impact their daily lives, causing symptoms like pain, swelling, fatigue, dizziness, and shortness of breath. The participants have tried |
| | various medications, but they often only provide temporary relief, leading to concerns about the risk of stroke and the high cost of medication, which can be as high as $32,000 a month. |
| Medical Condition | Dealing with Reflex Sympathetic Dystrophy (RSD), a chronic pain condition affecting the nervous system, was a challenging experience. Severe pain, swelling, and limited mobility in the knee were not adequately addressed by the initial doctor, requiring a second opinion. A specialist confirmed the diagnosis, but treatment was complicated by the workers' compensation system, limiting access to care and medication. |
| Mental Health | Participants have experienced various mental health issues, including anxiety, depression, PTSD, dissociative disorder, chronic depression, insomnia, and CPTSD, which have been exacerbated by difficulties in the healthcare system, stigma, and traumatic experiences. They have faced challenges such as feeling dismissed, long waitlists, uneducated providers, and inadequate care. Some have relied on self-researched coping mechanisms, while others have had to try multiple therapists and medications to find what works for them. |
| Healthcare Experience | The participants' experiences with the healthcare system are varied, with some reporting positive encounters and others facing challenges and frustrations. Common themes include:<br><br>1. **Lack of empathy and understanding**: Many participants felt disrespected, dismissed, or not taken seriously by healthcare providers, leading to feelings of frustration and mistrust.<br>2. **Difficulty navigating the system**: Participants often struggled to find the right doctors, get referrals, and access necessary treatments, highlighting the complexity and fragmentation of the healthcare system. |
| | 3. **Challenges with insurance and affordability**: Several participants faced issues with insurance coverage, medication costs, and affordability, which affected their ability to access necessary care.<br><br>4. **Importance of self-advocacy**: Many participants emphasized the need to take an active role in their own healthcare, researching their conditions, questioning doctors, and seeking second opinions to ensure they receive the best care possible.<br><br>5. **Positive experiences with supportive providers**: Some participants reported positive experiences with healthcare providers who were empathetic, knowledgeable, and willing to work with them to manage their conditions.<br><br>6. **Challenges with chronic conditions**: Participants with chronic conditions, such as diabetes, cancer, and mental health issues, often faced difficulties in managing their conditions and accessing necessary treatments.<br><br>7. **Impact of stigma and bias**: Some participants reported experiencing stigma and bias due to their conditions, such as HIV, addiction, or sickle cell disease, which affected their interactions with healthcare providers.<br><br>8. **Importance of support systems**: Participants often highlighted the importance of having a support system, including family, friends, and support groups, to help navigate the healthcare system and manage their conditions.<br><br>Overall, the participants' experiences highlight the need for a more patient-centered, empathetic, and supportive healthcare system that addresses the complex needs of individuals with various conditions. |
| Personal Experience | Living with chronic conditions and navigating the healthcare system can be a challenging and isolating experience. Participants have had to deal with severe side effects, stigma, and discrimination, while also learning to |
| | advocate for themselves and others. Despite these challenges, many have found ways to take care of themselves, such as eating healthy, researching, and trying new things. Supportive networks, including family and friends, have also played a crucial role in their journeys. Participants have learned to appreciate the preciousness of life and the importance of living in the moment, and have found ways to grow and self-discover through their experiences. |
| Healthcare System | The participants' experiences with the healthcare system are overwhelmingly negative, with many expressing frustration, disappointment, and feelings of neglect. Common themes include: <br><br> 1. Lack of support and understanding from healthcare providers, particularly for chronic conditions, mental health, and minority communities. <br> 2. Inadequate communication, empathy, and compassion from healthcare providers. <br> 3. Difficulty navigating the healthcare system, including accessing specialized care, transportation, and services. <br> 4. Insufficient education and awareness about various health conditions, including HIV, AIDS, and rare diseases. <br> 5. Perceived bias and discrimination within the healthcare system, particularly towards minority communities and women. <br> 6. Need for more emphasis on patient empowerment, self-advocacy, and personalized care. <br> 7. Importance of addressing the root causes of health problems, rather than just treating symptoms. <br> 8. Frustration with the lack of research and funding for certain health conditions. <br> 9. Difficulty accessing affordable care and medication, particularly for those |

|  | with limited resources or support. |
|  | 10. Need for better standards and regulations for caregiving companies and support for caregivers and their clients. |
|  | Overall, the participants' experiences highlight the need for a more patient-centered, compassionate, and equitable healthcare system that addresses the unique needs and challenges of diverse populations. |
| Caregiving Experience | Caregivers face numerous challenges, including lack of support, financial struggles, and emotional toll. Many participants expressed the importance of compassion, patience, and affection when caring for loved ones, particularly those with dementia or Alzheimer's. Some participants had to navigate complex healthcare systems, advocate for their loved ones' needs, and deal with the emotional impact of caregiving. Others appreciated the support of family members, healthcare providers, and online support groups. Several participants highlighted the need for respite care, financial assistance, and guidance for caregivers. Some also emphasized the importance of acknowledging stress and seeking help. Additionally, participants noted that healthcare providers should advise patients on healthy living and provide resources for caregivers. Overall, caregiving experiences varied, but many participants shared the common theme of needing more support and resources to effectively care for their loved ones. |
| Symptoms and Hospitalization | Initial misdiagnosis of nasal congestion led to hospitalization due to breathing difficulties and chest tightness. A series of tests, including blood tests and a kidney biopsy, resulted in a diagnosis of lupus. Repeated hospital visits were necessary to drain fluid from lungs and manage symptoms. |